\documentclass[pra,showpacs]{revtex4}
\usepackage{amssymb}
\usepackage{amsmath}
\usepackage{graphicx}
\usepackage{makeidx}
\usepackage{subfigure}

\begin{document}

\title{Suppression of the quantum-mechanical collapse by repulsive
interactions in a quantum gas}
\author{Hidetsugu Sakaguchi$^{1}$ and Boris A. Malomed$^{2}$}
\affiliation{$^{1}$Department of Applied Science for Electronics and Materials,
Interdisciplinary Graduate School of Engineering Sciences, Kyushu
University, Kasuga, Fukuoka 816-8580, Japan\\
$^{2}$Department of Physical Electronics, School of Electrical Engineering,
Faculty of Engineering, Tel Aviv University, Tel Aviv 69978, Israel}

\begin{abstract}
The quantum-mechanical collapse (alias \textit{fall onto the center } of
particles attracted by potential $-r^{-2}$) is a well-known issue in the
quantum theory. It is closely related to the \textit{quantum anomaly}, i.e.,
breaking of the scaling invariance of the respective Hamiltonian by the
quantization. We demonstrate that the mean-field repulsive nonlinearity
prevents the collapse and thus puts forward a solution to the
quantum-anomaly problem different from that previously developed in the
framework of the linear quantum-field theory. This solution may be realized
in the 3D or 2D gas of dipolar bosons attracted by a central charge, and in
the 2D gas of magnetic dipoles attracted by a current filament. In the 3D
setting, the dipole-dipole interactions are also taken into regard, in the
mean-field approximation, resulting in a redifinition of the scattering
length which accounts for the contact repulsion between the bosons. In lieu
of the collapse, the cubic nonlinearity creates a 3D ground state (GS),
which does not exist in the respective linear Schr\"{o}dinger equation. The
addition of the harmonic trap gives rise to a tristability, in the case when
the Schr\"{o}dinger equation still does not lead to the collapse. In the 2D
setting, the cubic nonlinearity is not strong enough to prevent the
collapse; however, the quintic term does it, creating the GS, as well as its
counterparts carrying the angular momentum (vorticity). Counter-intuitively,
such self-trapped 2D modes exist even in the case of a weakly \emph{%
repulsive} potential $r^{-2}$. The 2D vortical modes avoid the phase
singularity at the pivot ($r=0$) by having the amplitude diverging at $%
r\rightarrow 0$, instead of the usual situation with the amplitude of the
vortical mode vanishing at $r\rightarrow 0$ (the norm of the mode converges
despite of the singularity of the amplitude at $r\rightarrow 0$). In the
presence of the harmonic trap, the 2D quintic model with a weakly repulsive
central potential $r^{-2}$ gives rise to three confined modes, the middle
one being unstable, spontaneously developing into a breather. In both the 3D
and 2D cases, the GS wave functions are found in a numerical form, and also
in the form of an analytical approximation, which is asymptotically exact in
the limit of the large norm.
\end{abstract}

\maketitle

\section{Introduction and the physical settings}

It is well known that attractive potential $U(r)=-\left( U_{0}/2\right)
r^{-2}$ plays a critical role in quantum mechanics. Indeed, while the
corresponding Hamiltonian, taken in the scaled form, $\hat{H}=-\left(
1/2\right) \nabla ^{2}+U(r)$, obeys the scaling invariance,
\begin{equation}
\hat{H}(\alpha r)=H(r)/\alpha ^{2},  \label{H}
\end{equation}%
the quantization breaks the invariance, which is known as the \textit{%
quantum anomaly} , alias ``dimensional transmutation" \cite%
{anomaly}. A manifestation of the anomaly is that, at $U_{0}>$ $\left(
U_{0}\right) _{\mathrm{cr}}^{\mathrm{(3D)}}\equiv 1/4$, the corresponding 3D
(three-dimensional) Schr\"{o}dinger equation, $i\psi _{t}=\hat{H}\psi ,$
does not produce the ground state (GS); instead, it gives rise to the
quantum collapse, also known as \textit{fall onto the center} \cite{LL}. In
the 2D space, the critical value of the potential's strength is exactly
zero, $\left( U_{0}\right) _{\mathrm{cr}}^{\mathrm{(2D)}}=0$, and in the 1D
setting the same potential gives rise to a still stronger \textit{\
superselection} effect, effectively splitting the 1D space into two disjoint
subspaces, $x\gtrless 0$ \cite{superselection}.

A solution to the quantum-anomaly problem in the 3D case was proposed
outside of the realm of quantum mechanics, in terms of a linear
quantum-field-theory description, which introduces a renormalization and
imposes a GS with an \emph{arbitrary} spatial scale \cite{anomaly}. The
purpose of the present work is to demonstrate that a different solution to
this problem can be obtained within the framework of the mean-field
approach, taking into account the effective repulsive nonlinearity induced
by collisions of particles trapped in the potential under the consideration.
In the 3D setting, the usual cubic self-repulsive term will be sufficient
for this, while in the 2D geometry a stronger quintic term is necessary. As
a result, we will find the GS with a scale \emph{uniquely defined} by the
physical parameters. It will be demonstrated that the same nonlinearity is
helpful too below the critical point, where it makes the GS normalizable [at
$U_{0}<\left( U_{0}\right) _{\mathrm{cr}}^{\mathrm{(3D,2D)}}$, the linear
Schr\"{o}dinger equation gives rise to GS wave functions with the divergent
norm, see below].

The linear Schr\"{o}dinger equation with the critical potential is%
\begin{equation}
i\psi _{t}=-\frac{1}{2}\left( \nabla ^{2}+U_{0}r^{-2}-\Omega
^{2}r^{2}\right) \psi ,  \label{Sch0}
\end{equation}%
where the external trapping potential, $\left( \Omega ^{2}/2\right) r^{2}$,
is included too \cite{trap}. A physical realization of Eq. (\ref{Sch0}) in
the 3D space is provided by molecules with a permanent electric dipole
moment, $d$, interacting with charge $Q$ placed at the origin, which creates
electric field $\mathbf{E}=Q\mathbf{r}/r^{3}$. As demonstrated in a recent
experiment, the charged particle (ion) immersed into an ultracold gas may be
kept at the central position by means of the laser-trapping technique \cite%
{ion}. Assuming that the orientation of the dipole is locked to the local
field, i.e., $\mathbf{d}=\mathrm{sgn}(Q)d\left( \mathbf{r}/r\right) $, the
respective interaction potential is $U(r)=-\mathbf{d}\cdot \mathbf{E}$,
which corresponds to Eq. (\ref{Sch0}) with $U_{0}=2|Q|d$. This realization
is relevant to the capture of electrons by dipolar molecules, which was
studied in detail experimentally \cite{experiment,anomaly2}, and to
Bose-Einstein condensates (BECs) formed by dipolar molecules, such as Li-Cs
\cite{LiCs}. We stress that we consider the gas of permanent dipoles, while,
in the case of polarizable molecules, with $\mathbf{d}=\varepsilon \mathbf{E}
$, where $\varepsilon $ is the polarizability, the attractive potential is
different, $U=-\varepsilon Q^{2}r^{-4}$.

If the gas of ultracold dipolar molecules is trapped in a pancake-shaped
configuration sustained by an appropriate external potential \cite{2D-review}%
, inserting the central electric charge provides for the realization of the
2D version of Eq. (\ref{Sch0}). An additional realization of the 2D setting
is offered by a gas of atoms (chromium \cite{Cr,France}) or molecules (such
as $^{87}$Rb$_{2}$ \cite{Rb}) carrying magnetic moments, the attractive
potential being induced by the magnetic field of a transverse current
filament, with the orientation of the dipoles locked to the local magnetic
field. Actually, the orientations of the dipoles may form the configurations
obeying Eq. (\ref{Sch0}) in the dipolar BEC produced by means of the
all-optical trapping \cite{France}, which does not freeze the dipoles into
the confining magnetic field.

In the mean-field approximation, the contact repulsive interaction in the
bosonic gas is represented by the cubic term \cite{Pit}. With the addition
of this term, the linear Schr\"{o}dinger equation (\ref{Sch0}) is replaced
by the Gross-Pitaevskii equation (GPE),
\begin{equation}
i\psi _{t}=-\left( 1/2\right) \left( \nabla ^{2}+U_{0}r^{-2}-\Omega
^{2}r^{2}\right) \psi +\left\vert \psi \right\vert ^{2}\psi .  \label{GPE}
\end{equation}%
The relation between the scaled variables and constants, in terms of which
Eq. (\ref{GPE}) is written, and their counterparts defined in the usual
physical units \cite{Pit} is%
\begin{equation}
\mathbf{r}=\frac{\mathbf{r}_{\mathrm{ph}}}{r_{0}},~t=\frac{\hbar }{mr_{0}^{2}%
}t_{\mathrm{ph}},~~\psi =2\sqrt{\pi a_{s}}r_{0}\psi _{\mathrm{ph}},~U_{0}=%
\frac{m}{\hbar ^{2}}\left( U_{0}\right) _{\mathrm{ph}},~\Omega =\frac{%
mr_{0}^{2}}{\hbar }\Omega _{\mathrm{ph}},  \label{ph}
\end{equation}%
where $m$ and $a_{s}$ are bosonic mass and \textit{s}-scattering length,
which accounts for the repulsive interactions, and $r_{0}$ is an arbitrary
spatial scale. Accordingly, the total number of bosons in the gas is given by%
\begin{equation}
N_{\mathrm{ph}}=\int \left\vert \psi _{\mathrm{ph}}(\mathbf{r}_{\mathrm{ph}%
})\right\vert ^{2}d\mathbf{r}_{\mathrm{ph}}\equiv \frac{r_{0}N}{4\pi a_{s}},
\label{Nph}
\end{equation}%
where $N=\int \left\vert \psi (\mathbf{r})\right\vert ^{2}d\mathbf{r}$ is
the norm of the scaled wave function.

In the form of Eq. (\ref{GPE}), the GPE neglects the dipole-dipole
interactions between the particles. However, these can be readily taken into
account, using the same mean-field approximation which is used to derive the
GPE. Indeed, the local density of the dipole moment in the gas (i.e., the
polarization of the medium) is $\mathbf{P}=\mathbf{d}\left\vert \psi (%
\mathbf{r})\right\vert ^{2}$, hence the additional electric field generated
by the polarization, $\mathbf{E}_{d}$, is determined by the Poisson
equation, $\nabla \cdot \left( \mathbf{E}_{d}+4\pi \mathbf{P}\right) =0$,
which yields $\mathbf{E}_{d}=-4\pi \mathbf{P}\equiv -4\pi \mathbf{d}%
\left\vert \psi (\mathbf{r})\right\vert ^{2}$ (this solution for $\mathbf{E}%
_{d}$ is definitely a unique one for the spherically symmetric
configurations considered below). Finally, the extra term in the GPE induced
by the interaction of the local dipole with the collective field, $\mathbf{E}%
_{d}$, created by all other dipoles is%
\begin{equation}
-\left( \mathbf{d}\cdot \mathbf{E}_{d}\right) \psi \equiv 4\pi
d^{2}\left\vert \psi \right\vert ^{2}\psi .  \label{dip}
\end{equation}%
Obviously, this term, if added to Eq. (\ref{GPE}), may be absorbed into a
redefinition of the effective scattering length accounting for the repulsion
between the particles. In the underlying physical units, this amounts to
\begin{equation}
a_{s}\rightarrow \left( a_{s}\right) _{\mathrm{eff}}\equiv
a_{s}+md^{2}/\hbar ^{2},  \label{eff}
\end{equation}%
where $m$ is the mass of the dipolar molecule.

Unlike the quantization, the inclusion of the nonlinearity does not break
the underlying scaling invariance of the Hamiltonian, cf. Eq. (\ref{H}).
Indeed, under the combined transformation,
\begin{equation}
\mathbf{r}\rightarrow \alpha \mathbf{r},\psi \rightarrow \alpha ^{-1}\psi
,\Omega \rightarrow \alpha ^{-2}\Omega .  \label{alpha}
\end{equation}%
the total energy of the condensate described by GPE (\ref{GPE}),%
\begin{equation}
E_{D}=\frac{1}{2}\int \left[ \left\vert \nabla \psi \right\vert ^{2}-\left(
U_{0}r^{-2}-\Omega ^{2}r^{2}\right) \left\vert \psi \right\vert
^{2}+\left\vert \psi ^{4}\right\vert \right] d\mathbf{r},  \label{E}
\end{equation}%
features scaling $E\rightarrow E/\alpha ^{4-D}$, where $D$ is the dimension (%
$3$ or $2$).

It is relevant to mention that a quantum anomaly was very recently predicted
in a model described by the GPE in the 2D space, for a harmonically trapped
gas of bosons interacting through the two-dimensional repulsive
delta-functional potential \cite{Olshanii}. The anomaly breaks the specific
scaling invariance of this gas, which holds in the mean-field approximation
\cite{Pit97}.

The rest of the paper is organized as follows. In Section II, we briefly
recapitulate the description of the 3D and 2D collapses in the framework of
quantum mechanics, extending it through the inclusion of self-similar
nonstationary solutions, in addition to the known stationary ones. Section
III reports the basic results obtained in the 3D case, which demonstrate the
creation of the previously missing GS by the self-repulsive cubic
nonlinearity at $U_{0}>1/4$, as well as making the GS normalizable at $%
U_{0}<1/4$. It is demonstrated too that, in latter case, the inclusion of
the harmonic trapping potential gives rise to a \textit{tristability} of
bound states. The results obtained in the 2D model with the quintic
repulsive term are reported in Section IV. In that case, the GS also
replaces the quantum-collapse regime for $U_{0}>0$. In the case of $%
0<-U_{0}<1/4$ (i.e., in the case of the \emph{weakly repulsive} potential),
the quintic nonlinearity also gives rise to the GS with a normalizable wave
function. In the latter case, three confined modes are found in the presence
of the harmonic trap, like in the 3D setting, but the middle mode is
unstable, spontaneously transforming itself into a breather. The paper is
concluded by Section V.

\section{The collapse in quantum mechanics}

In the framework of Eq. (\ref{Sch0}), 3D and 2D stationary states, with
angular quantum numbers $l$ and $m$, are looked for as
\begin{eqnarray}
\psi _{\mathrm{3D}} &=&\exp (-i\mu t)Y_{lm}\left( \theta ,\varphi \right)
\phi (r),  \label{psi} \\
\psi _{\mathrm{2D}} &=&\exp (-i\mu t+il\varphi )\phi (r),  \label{2D}
\end{eqnarray}%
where $\theta $ and $\varphi $ are the angular coordinates and $Y_{lm}\left(
\theta ,\varphi \right) $ is the spherical harmonic. Then, an \emph{exact}
3D wave function in the form of Eq. (\ref{psi}) can be readily found, as a
solution to Eq. (\ref{Sch0}), for $U_{l}\equiv U_{0}-l(l+1)<1/4$, cf. Ref.
\cite{trap}:
\begin{gather}
\phi (r)=\phi _{0}r^{-\sigma _{\pm }}\exp \left( -\Omega r^{2}/2\right) ,
\label{exact} \\
\mu =\Omega \left( \frac{3}{2}-\sigma _{\pm }\right) ,~\sigma _{\pm }\equiv
\frac{1}{2}\pm \sqrt{\frac{1}{4}-U_{l}},  \label{sigma3D}
\end{gather}%
the GS corresponding to the smaller value of $\mu $, i.e., $\sigma _{+}$.
This solution is also relevant for $U_{l}<0$ (the repulsive potential), with
the respective norm,
\begin{equation}
N_{\mathrm{3D}}=4\pi \int_{0}^{\infty }\phi ^{2}(r)r^{2}dr,  \label{N}
\end{equation}%
converging (for $\sigma _{+}$) if $U_{l}>-3/4$.

In two dimensions, the GS solution to Eq. (\ref{Sch0}) exist only for $%
U_{l}\equiv U_{0}-l^{2}<0$, in the exact form given by Eqs. (\ref{2D}) and (%
\ref{exact}), but with
\begin{equation}
\mu =\Omega \left( 1-\sigma _{\pm }\right) ,\sigma _{\pm }=\pm \sqrt{-U_{l}},
\label{sigma2D}
\end{equation}%
cf. Eq. (\ref{sigma3D}). Like in the 3D case, the GS corresponds to $\sigma
_{+}$, the 2D norm of this solution,
\begin{equation}
N_{\mathrm{2D}}=2\pi \int_{0}^{\infty }\phi ^{2}(r)rdr,  \label{N2D}
\end{equation}%
converging if $U_{l}>-1$.

\begin{figure}[t]
\begin{center}
\includegraphics[height=3.5cm]{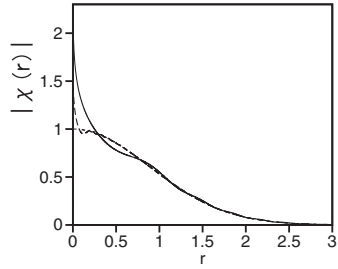}
\end{center}
\caption{Three radial profiles of $|\protect\chi (r,t)|=|\protect\psi %
(r)|r^{1/2}$ at $t=0$, $0.005$ and $0.1$ (dotted, dashed, and solid curves),
as produced by the numerical solution of the radial version of Eq. (\protect
\ref{Sch0}) in 3D with $U_{0}=0.27$ and $\Omega ^{2}=0.1$. Note that this
value of $U_{0}$ slightly exceeds the critical one, $\left( U_{0}\right) _{%
\mathrm{cr}}^{\mathrm{(3D)}}=1/4$.}
\label{fig1}
\end{figure}

Past the critical point, i.e., for $U_{l}>1/4$ in 3D, and for any $U_{l}>0$
in 2D, the asymptotic form of solutions for $\phi (r)$ at $r\rightarrow 0$
is \cite{LL}
\begin{equation}
\phi (r)\approx \phi _{0}\left\{
\begin{array}{c}
r^{-1/2}\cos \left( \sqrt{U_{l}-1/4}\ln \left( r/r_{0}\right) \right) ,~D=3,
\\
\cos \left( \sqrt{U_{l}}\ln \left( r/r_{0}\right) \right) ,~D=2,%
\end{array}%
\right.   \label{mu=0}
\end{equation}%
with arbitrary constants $\phi _{0}$ and $r_{0}$ (these stationary solutions
are exact but unnormalizable ones for $\mu =\Omega =0$). The infinite number
of radial nodes (zeros) in solutions (\ref{mu=0}) implies the nonexistence
of the GS in these cases, and represents the phenomenon of the quantum
anomaly. Accordingly, the nonstationary wave function obeying Eq. (\ref{Sch0}%
) is expected to collapse, eventually. The latter conjecture was confirmed
by direct simulations. As an example, in Fig. 1 we display results of the
simulations of the 3D spherically symmetric solution, with $l=m=0$. The
simulations were performed by means of the split-step Fourier method using $%
2^{16}$ modes. As the initial condition, we took $\psi (r)=r^{-1/2}\exp
(-\Omega r^{2}/2)$, which is the exact stationary wave function for $%
U_{0}=1/4$, i.e., the one at the critical point, see Eqs. (\ref{exact}) and (%
\ref{sigma3D}). The snap-shot profiles of $r^{1/2}|\psi (r)|$ shown in Fig. %
\ref{fig1} testify to the rapid growth of the amplitude of the solution at $%
r=0$, which is a signature of the development of the collapse. The growth
eventually ceases in the simulations, due to the finite mesh size of the
numerical scheme.

A more general version of asymptotic solution (\ref{mu=0}) can be obtained
by means of substitution
\begin{equation}
\psi \left( r,t\right) =r^{-1/2}Y_{lm}\left( \theta ,\varphi \right) \chi
\left( r,t\right) ,  \label{psichi}
\end{equation}%
which makes Eq. (\ref{Sch0}) for the 3D time-dependent wave function
equivalent to its 2D counterpart:
\begin{equation}
i\chi _{t}=-\left( 1/2\right) \left[ \partial _{r}^{2}+r^{-1}\partial
_{r}+\left( U_{l}-1/4\right) r^{-2}-\Omega ^{2}r^{2}\right] \chi .
\label{linchi}
\end{equation}%
Then, omitting the term $\sim \Omega ^{2}$, time-dependent solutions to Eq. (%
\ref{linchi}) may be looked for in a self-similar form, $\chi \left(
r,t\right) =\chi \left( \Lambda \right) $, where%
\begin{equation}
\Lambda \equiv \ln \left( \left( \rho _{0}^{2}/r^{2}\right) \left( t/\tau
_{0}\right) \right) ,  \label{Lambda}
\end{equation}%
with arbitrary radial and temporal scales, $\rho _{0}$ and $\tau _{0}$. This
substitution transforms Eq. (\ref{linchi}) with $\Omega =0$ into an ordinary
differential equation,
\begin{equation}
ie^{-\Lambda }\frac{d\chi }{d\Lambda }=-\frac{\tau _{0}}{2\rho _{0}^{2}}%
\left[ 4\frac{d^{2}\chi }{d\Lambda ^{2}}+\left( U_{l}-\frac{1}{4}\right)
\chi \right] .  \label{ODE}
\end{equation}%
Explicit solutions to Eq. (\ref{ODE}) can be found in an asymptotic form,
corresponding to $\Lambda \rightarrow \infty $ (i.e., $r\rightarrow 0$
and/or $t\rightarrow \infty $), which is a spatiotemporal generalization of
the stationary solution (\ref{mu=0}):%
\begin{eqnarray}
\chi  &=&\phi _{0}\cos \left( (1/2)\sqrt{U_{l}-1/4}\Lambda \right)   \notag
\\
&&+i\phi _{0}\frac{\sqrt{U_{l}-1/4}}{4U_{l}+3}\frac{r^{2}}{t}\left[ \sin
\left( (1/2)\sqrt{U_{l}-1/4}\Lambda \right) \right.   \notag \\
&&\left. +\sqrt{U_{l}-1/4}\cos \left( (1/2\sqrt{U_{l}-1/4}\Lambda \right) %
\right] ,  \label{self-similar}
\end{eqnarray}%
the second term being a small correction to the first one. Through
substitution (\ref{psichi}), solution (\ref{self-similar}) pertains to the
3D case, and, on the other hand, expression (\ref{self-similar}) with $\sqrt{%
U_{l}-1/4}$ replaced by $\sqrt{U_{l}}$ directly applies to the 2D situation.

With regard to Eq. (\ref{Lambda}), the meaning of solution (\ref%
{self-similar}) in both the 3D and 2D cases may be understood by noting that
the number of nodes of the wave function [i.e., the number of the
corresponding zeros of $\cos \left( (1/2)\sqrt{U_{l}-1/4}\Lambda \right) $]
which pass a point with fixed $r$ grows with time as $n\approx \left[ \sqrt{%
U_{l}-1/4}/\left( 2\pi \right) \right] \left[ \ln \left(
r_{0}^{2}/r^{2}\right) +\ln \left( t/t_{0}\right) \right] $. This actually
implies the development of the quantum-anomaly structure in time, which
expands outwards.

\section{The 3D ground state created by the cubic self-repulsive nonlinearity%
}

\subsection{The ground state in the absence of the external trap}

Spherically symmetric stationary solutions to the GPE in the form of Eq. (%
\ref{GPE}) are looked for as
\begin{equation}
\psi \left( r,t\right) =e^{-i\mu t}r^{-1}\chi _{\mathrm{3D}}(r),
\label{psichi3D}
\end{equation}%
with real function $\chi $ obeying equation
\begin{equation}
\mu \chi _{\mathrm{3D}}=-\frac{1}{2}\left[ \chi _{\mathrm{3D}}^{\prime
\prime }+\left( U_{0}r^{-2}-\Omega ^{2}r^{2}\right) \chi _{\mathrm{3D}}%
\right] +r^{-2}\chi _{\mathrm{3D}}^{3}.  \label{chi}
\end{equation}%
The expansion of solutions to Eq. (\ref{chi}) at $r\rightarrow 0$ is
\begin{equation}
\chi _{\mathrm{3D}}(r)=\sqrt{U_{0}/2}+\chi _{1}r^{s/2},~s=1+\sqrt{1+8U_{0}},
\label{r=0}
\end{equation}%
where $\chi _{1}$ is a free constant. For \emph{any} $U_{0}>0$, the cubic
nonlinearity supports finite-norm states without the help of the external
trapping potential ($\Omega =0$). Indeed, in this case the asymptotic form
of the solution with $\mu <0$ at $r\rightarrow \infty $ is $\chi _{\mathrm{3D%
}}=\chi _{0}\exp \left( -\sqrt{-2\mu }r\right) $, with some constant $\chi
_{0}$. Combining it with the asymptotic form (\ref{r=0}) valid at $%
r\rightarrow 0$, one may use, as the simplest analytical approximation, the
following interpolation for the 3D modes which represent the GS with given
norm $N$:{\
\begin{equation}
\psi _{\mathrm{3D}}^{(\Omega =0)}=\sqrt{\frac{U_{0}}{2}}e^{-i\mu t}r^{-1}e^{-%
\sqrt{-2\mu }r},~\mu =-\frac{1}{2}\left( \frac{\pi U_{0}}{N_{\mathrm{3D}}}%
\right) ^{2}.  \label{inter}
\end{equation}%
} The small term $\chi _{1}r^{s/2}$ from Eq. (\ref{r=0}) is not included
into this approximation, and the norm was calculated by the substitution of
analytical expression (\ref{inter}) into integral (\ref{N}). Actually, in
the limit of $\mu \rightarrow -0$, Eq. (\ref{inter}) gives an asymptotically
exact solution (rather than being simply an interpolation), and its limit
form corresponding to $\mu =0$ and $N=\infty $ is an exact (although
unnormalizable) solution.

\begin{figure}[t]
\begin{center}
\includegraphics[height=3.5cm]{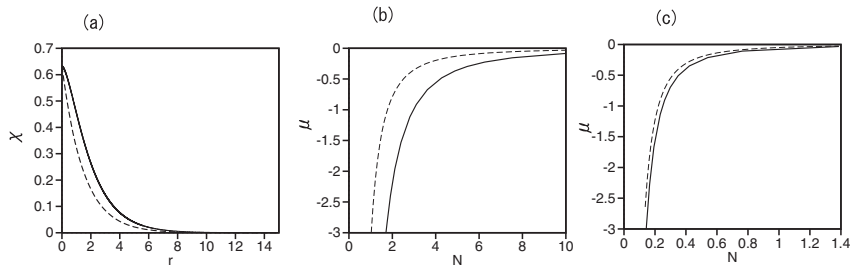}
\end{center}
\caption{ (a) A typical example of the ground state, shown in terms of $%
\protect\chi (r)\equiv r\left\vert \protect\psi (r)\right\vert $, in the 3D
nonlinear model without the external trap ($\Omega =0$), for $U_{0}=0.8$ and
$\protect\mu =-0.225$. Panels (b) and (c) display curves $\protect\mu (N)$
for the ground-state families with $U_{0}=0.8$ and $0.1$, respectively. In
all the panels, the solid and dashed curves depict, severally, the numerical
results and analytical approximation (\protect\ref{inter}). In particular,
the latter predicts $N(\protect\mu =-0.225)=5.30$ for $U_{0}=0.8$ [the case
shown in (a)], while the numerically found counterpart of this value is $N_{%
\mathrm{num}}(\protect\mu =-0.225)=6.26$. The convergence of the numerical
and analytical curves for $N(\protect\mu )$ at $\protect\mu \rightarrow
-\infty $ corresponds to the fact that Eq. (\protect\ref{inter}) gives an
asymptotically exact solution in this limit.}
\label{fig2}
\end{figure}

Numerical solutions of Eq. (\ref{chi}) were found by means of the shooting
method. A typical example of the GS solution, along with the respective
approximation (\ref{inter}), is displayed in Fig. \ref{fig2}(a) for $%
U_{0}=0.8$, which is essentially \emph{larger} than the critical attraction
strength, $\left( U_{0}\right) _{\mathrm{cr}}^{\mathrm{(3D)}}=1/4$, beyond
which the GS does not exist in the framework of the linear Schr\"{o}dinger
equation (\ref{Sch0}). Further, Figs. \ref{fig2}(b) and \ref{fig2}(c)
represent the entire family of the solutions for two values, $U_{0}=0.8$ and
$0.1$, which are, respectively, larger and smaller than $1/4$. The
conclusion is that, in the nonlinear model, the GS exists for \emph{all}
values of $U_{0}$ and $N$. Thus, the self-repulsive cubic term completely
suppresses the quantum collapse in the 3D space and offers an alternative
solution to the quantum-anomaly problem \cite{anomaly}, by creating the GS
where it does not exist in the linear Schr\"{o}dinger equation.

The analytical approximation (\ref{inter}) suggests an estimate for the
radial size of the GS created by the repulsive nonlinearity:%
\begin{equation}
R_{\mathrm{GS}}^{\mathrm{(3D)}}\equiv \frac{4\pi }{N_{\mathrm{3D}}}%
\int_{0}^{\infty }\left\vert \psi _{\mathrm{3D}}^{(\Omega =0)}(r)\right\vert
^{2}r^{3}dr=\frac{N_{\mathrm{3D}}}{2\pi U_{0}}.  \label{R}
\end{equation}%
It is relevant to rewrite this estimate in terms of the physical units, as
per Eqs. (\ref{ph}), (\ref{Nph}), and (\ref{eff}):%
\begin{equation}
\left( R_{\mathrm{GS}}^{\mathrm{(3D)}}\right) _{\mathrm{ph}}\equiv r_{0}R_{%
\mathrm{GS}}=\frac{2\left( \hbar ^{2}a_{s}+md^{2}\right) N_{\mathrm{ph}}}{%
m\left( U_{0}\right) _{\mathrm{ph}}},  \label{Rph}
\end{equation}%
which gives the radius of the GS in terms of the physical parameters of the
model. Note that arbitrary spatial scale $r_{0}$, which was used in
rescalings (\ref{ph}) and (\ref{Nph}), \emph{does not} appear in Eq. (\ref%
{Nph}) (actually, it cancels out). It is natural that $R_{\mathrm{GS}}$
shrinks to zero at $N\rightarrow 0$, which implies the transition to the
collapse in the framework of the linear Schr\"{o}dinger equation.

The existence of the GS with the \emph{finite norm} at $U_{0}<1/4$ in the
nonlinear model with $\Omega =0$ is worthy noting too, as the corresponding
stationary solutions to the linear equation, see Eqs. (\ref{psi}) and (\ref%
{exact}), are unnormalizable for $\Omega =0$. Lastly, simulations of the
radial version of Eq. (\ref{GPE}) with arbitrary random perturbations added
to the stationary solutions (not shown here) demonstrate that the entire GS
family is stable. Those simulations did not test the stability of the GS
against tesseral (spherical-symmetry-breaking) perturbations, but the
repulsive character of both the contact and dipole-dipole interactions (in
the mean-field approximation) makes the presence of a symmetry-breaking
instability implausible.

It is relevant to mention that, as follows from Eq. (\ref{r=0}), the
(scaled) energy of the GS, calculated, as per definition (\ref{E}), in a
regularized form, i.e., for $r\geq \varrho \rightarrow 0$, contains a
diverging term,
\begin{equation}
\tilde{E}_{\mathrm{3D}}=\pi U_{0}\left( 2-U_{0}\right) \left( 2\varrho
\right) ^{-1},  \label{diverging}
\end{equation}%
which may be removed by means of the renormalization, cf. Refs. \cite%
{anomaly,anomaly2}. The vanishing of $\tilde{E}_{\mathrm{3D}}$ at $U_{0}=2$
seems to be a formal peculiarity, rather than a physical feature of the
model.

\subsection{Effects of the harmonic trap: the tristability}

\begin{figure}[t]
\begin{center}
\includegraphics[height=4cm]{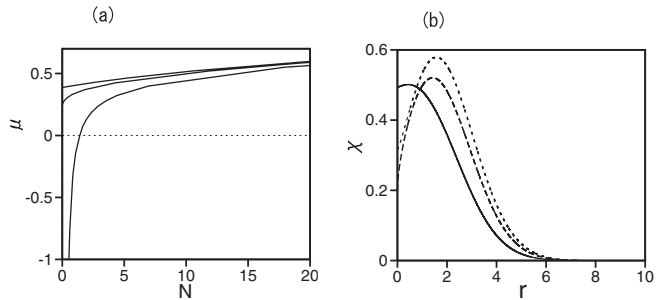}
\end{center}
\caption{(a) Curves $\protect\mu (N)$ for three families of the 3D confined
modes at $\Omega ^{2}=0.1$ and $U_{0}=0.2$. (b) Profiles of typical modes
belonging to the upper, middle, and lower branches in (a) are shown by the
solid, dashed, and dotted curves, respectively. In panel (b), $\protect\chi %
\equiv r\left\vert \protect\psi (r)\right\vert $ for the solution belonging
to the lower branch in (a), while $\protect\chi \equiv r^{\protect\sigma %
_{\pm }}\left\vert \protect\psi (r)\right\vert $ for the two upper ones,
with the same $\protect\sigma _{\pm }$ as in Eq. (\protect\ref{sigma3D}).
The profiles are shown in (b) for $N=10.2$, the respective values of $%
\protect\mu $ being $0.517$, $0.503$, and $0.460$.}
\label{fig3}
\end{figure}

The addition of the harmonic trap deforms the GS family reported above for $%
\Omega =0$. As shown in Fig. \ref{fig3}, at $U_{0}<1/4$ the nonlinear model
with $\Omega >0$ supports two additional families of 3D confined modes, thus
featuring a \textit{tristability}. The lowest curve in Fig. \ref{fig3}(a)
represents the deformed GS branch produced by the numerical solution of Eq.~(%
\ref{chi}), while two upper branches represent a nonlinear deformation of
the exact solutions generated by linear equation (\ref{Sch0}) in the form of
Eqs. (\ref{psi}), (\ref{exact}), and (\ref{sigma3D}) (note that in the
linear limit, i.e., at $N\rightarrow 0$, the GS created by the nonlinearity
disappears by falling to $\mu \rightarrow -\infty $). The upper branches
merge and disappear at $U_{0}\rightarrow 1/4$. The stability of all the
three families was verified by direct simulations of the radial version of
Eq. (\ref{GPE}) (not shown here). In the limit of large $N$, all the three
branches in Fig. \ref{fig3}(a) asymptotically approach an expression
predicted by the Thomas-Fermi approximation,
\begin{equation}
\left( N_{\mathrm{TF}}\right) _{\mathrm{3D}}\approx \left( 16\sqrt{2}\pi
/15\Omega ^{3}\right) \mu ^{5/2}.  \label{TF3D}
\end{equation}%

\subsection{3D vortical modes}

\begin{figure}[t]
\begin{center}
\includegraphics[height=3.5cm]{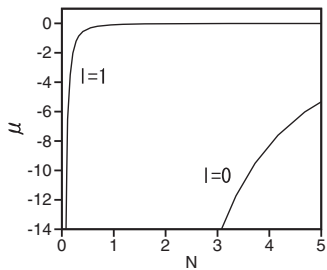}
\end{center}
\caption{The $\protect\mu (N)$ curves at $\Omega ^{2}=0$ and $U_{0}=2.2$ for
3D modes with angular quantum numbers $\left( l,m\right) =(0,0)$ and $\left(
1,0\right) $, obtained by means of the approximate ansatz (\protect\ref{Y}).}
\label{fig4}
\end{figure}

Nonlinear solutions for 3D states carrying the angular momentum may be
considered by means of the averaging in the angular variables, $\left\vert
Y_{lm}\left( \theta ,\varphi \right) \right\vert ^{2}Y_{lm}\left( \theta
,\varphi \right) \rightarrow c_{lm}Y_{lm}\left( \theta ,\varphi \right) $
with $c_{lm}=\int_{0}^{\pi }\sin \theta d\theta \int_{0}^{2\pi }\left\vert
Y_{lm}\left( \theta ,\varphi \right) \right\vert ^{4}d\varphi $ (in
particular, $c_{10}=9/5$, $c_{11}=6/5$). In this approximation, ansatz
\begin{equation}
\psi =c_{lm}^{-1/2}Y_{lm}\left( \theta ,\varphi \right) e^{-i\mu t}r^{-1}%
\tilde{\chi}(r),  \label{Y}
\end{equation}%
cf. Eq. (\ref{psichi3D}), leads to Eq. (\ref{chi}) with $\chi $ replaced by $%
\tilde{\chi}$, and $U_{0}$ substituted by $U_{0}-l\left( l+1\right) $. In
Fig. \ref{fig4} we compare numerically found curves $\mu (N)$ for the modes
with $(l,m)=(0,0)$ and $(1,0)$, fixing $U_{0}=2.2$ and $\Omega =0$. In this
figure, the mode corresponding to $(l,m)=(1,0)$ was generated using the
spherically symmetric equation, with the norm modified as per the averaging
approximation, see Eq. (\ref{Y}).

\section{2D ground states created by the quintic self-repulsive nonlinearity}

\subsection{The ground state in the absence of the external trap}

As said above, the GPE in the form of Eq. (\ref{GPE}) is relevant in the 2D
case too. However, the 2D norm of the solution with asymptotic form $\sim
r^{-1}$ at $r\rightarrow 0$, which follows from this equation [see Eq. (\ref%
{inter})], diverges. In other words, the cubic self-repulsion is not strong
enough to prevent the collapse in the 2D setting. On the other hand, the GPE
may also include the quintic repulsive term accounting for three-body
collisions, provided that the collisions do not give rise to conspicuous
losses \cite{3-body}. The 2D axisymmetric GPE with the dominating quintic
term is
\begin{equation}
i\psi _{t}=-\left( 1/2\right) \left( \psi _{rr}+r^{-1}\psi
_{r}+U_{l}r^{-2}-\Omega ^{2}r^{2}\right) \psi +\left\vert \psi \right\vert
^{4}\psi ,  \label{psi2d}
\end{equation}%
[recall $U_{l}\equiv U_{0}-l^{2}$, if vorticity $l$ is present, see Eq. (\ref%
{2D}); unlike the 3D case, the vorticity may be kept in the 2D
analysis based on radial equations]. In Eq. (\ref{psi2d}), we
neglect inessential cubic terms, including the one which may be
generated by the dipole-dipole interactions, as in Eq. (\ref{eff}).
If the cubic terms are kept, they do not significantly affect the
results presented below [in particular, they do not alter the first
term in expansion (\ref{r=0-2D}) at $r\rightarrow 0$].

The total energy of the BEC described by 2D equation (\ref{psi2d}) is%
\begin{equation}
E_{\mathrm{2D}}=\frac{1}{2}\int_{0}^{\infty }\left[ \left\vert \nabla \psi
\right\vert ^{2}-\left( U_{0}r^{-2}-\Omega ^{2}r^{2}\right) \left\vert \psi
\right\vert ^{2}+\frac{2}{3}\left\vert \psi ^{6}\right\vert \right] d\mathbf{%
r},  \label{E2D}
\end{equation}%
cf. Eq. (\ref{E}). This energy features the invariance with respect to the
scaling transformation, $E_{\mathrm{2D}}\rightarrow E_{\mathrm{2D}}/\alpha $%
, with the difference from the 3D case in that the wave function is
transformed as per $\psi \rightarrow \alpha ^{-1/2}\psi $, cf. Eq. (\ref%
{alpha}).

Stationary solutions to Eq. (\ref{psi2d}) are looked for as%
\begin{equation}
\psi \left( r,t\right) =e^{-i\mu t}r^{-1/2}\chi _{\mathrm{2D}}(r),
\label{psichi2D}
\end{equation}%
cf. Eq. (\ref{psichi3D}), which yields an equation for $\chi (r)$:%
\begin{equation}
\mu \chi _{\mathrm{2D}}=-\frac{1}{2}\left[ \chi _{\mathrm{2D}}^{\prime
\prime }+\left( U_{l}+\frac{1}{4}\right) r^{-2}\chi \right] +r^{-2}\chi ^{5},
\label{chi2D}
\end{equation}%
where we set $\Omega =0$, cf. Eq. (\ref{chi}) in 3D. The expansion of the
solution to Eq. (\ref{chi2D}) at $r\rightarrow 0$ is
\begin{equation}
\chi =\left[ \frac{1}{2}\left( U_{l}+\frac{1}{4}\right) \right] ^{1/4}+\chi
_{1}r^{s},  \label{r=0-2D}
\end{equation}%
where $s=\left( 1/2\right) \left( 1+\sqrt{5+16U_{l}}\right) $, and $\chi _{1}
$ is an arbitrary constant, cf. Eq. (\ref{r=0}) in the 3D case. The solution
with a finite norm exists at $U_{l}>-1/4$, representing, at $U_{l}>0$, the
suppression of the collapse and creation of the GS by the quintic
nonlinearity.

As concerns the vorticity, in the usual situation the amplitude of the
corresponding mode must vanish at point $r=0$, where the phase cannot be
defined. However, in the present case the solution features a different
solution to the phase-singularity problem: instead of vanishing, the
amplitude \emph{diverges} at $r\rightarrow 0$ -- as $\left\vert \psi
(r)\right\vert \approx \left[ \left( 1/2\right) \left( U_{l}+1/4\right) %
\right] ^{1/4}r^{-1/2}$, according to Eqs. (\ref{psichi2D}) and (\ref{r=0-2D}%
). This possibility is akin to the fact that the Bessel equation for
amplitudes of vortex modes with topological charge $l$ gives rise to two
solutions, $J_{\pm |l|}\left( \mathrm{const}\cdot r\right) $, the one with
index $-|l|$ being usually omitted as a singular one. However, in the
present case the singular solution is acceptable, as it provides for the
convergence of 2D norm (\ref{N2D}).

The energy of the GS, if calculated at $r\geq \varrho \rightarrow 0$,
contains a divergent term, $\tilde{E}_{\mathrm{2D}}=\pi \sqrt{U_{l}+1/4}%
\left( 1-2U_{l}\right) \left( 3\sqrt{2}\varrho \right) ^{-1}$, cf. the
similar term (\ref{diverging}) in the 3D setting. This term, which may be
removed by the renormalization procedure, and its vanishing at $U_{l}=1/2$,
do not seem to be physically significant features of the model.

Combining the 2D asymptotic form (\ref{r=0-2D}), valid at $r\rightarrow 0$,
and its counterpart at $r\rightarrow \infty $, $\chi _{\mathrm{2D}}\approx
\chi _{0}\exp \left( -\sqrt{-2\mu }r\right) $, and making use of definition (%
\ref{N2D}) for the 2D norm, we obtain an analytical interpolation formula
for the GS family, in the absence of the external trap ($\Omega =0$):
\begin{eqnarray}
\psi _{\mathrm{2D}}^{(\Omega =0)} &=&\left[ \frac{1}{2}\left( U_{l}+\frac{1}{%
4}\right) \right] ^{1/4}e^{-i\mu t}r^{-1/2}e^{-\sqrt{-2\mu }r},  \notag \\
\mu  &=&-\left( U_{l}+\frac{1}{4}\right) \left( \frac{\pi }{2N_{\mathrm{2D}}}%
\right) ^{2},  \label{inter2D}
\end{eqnarray}%
cf. approximation (\ref{inter}) in the 3D case. Similar to the situation in
the 3D case, Eq. (\ref{inter2D}) gives an asymptotically exact solution
(rather than a mere interpolation) for $\mu \rightarrow -0$, and an exact
solution with the infinite norm at $\mu =0$. The approximation (\ref{inter2D}%
) makes it possible to define the radial size of the two-dimensional GS
created by the quintic nonlinearity, cf. Eq. (\ref{R}) in the 3D case:%
\begin{equation}
R_{\mathrm{GS}}^{\mathrm{(2D)}}\equiv \frac{2\pi }{N_{\mathrm{2D}}}%
\int_{0}^{\infty }\left\vert \psi _{\mathrm{2D}}^{(\Omega =0)}(r)\right\vert
^{2}r^{2}dr=\frac{N_{\mathrm{2D}}}{\pi \sqrt{2\left( U_{0}+1/4\right) }}.
\end{equation}

Note that the quintic term supports the GS in 2D even at $0<-U_{l}<1/4$,
when the central potential is \emph{repulsive}. The correctness of this
counter-intuitive conclusion is corroborated by the above-mentioned fact
that the analytical approximation (\ref{inter2D}) gives the asymptotically
exact solution for $\mu \rightarrow 0$, \emph{including} the case of $%
0<-U_{l}<1/4$.

\begin{figure}[t]
\begin{center}
\includegraphics[height=3.5cm]{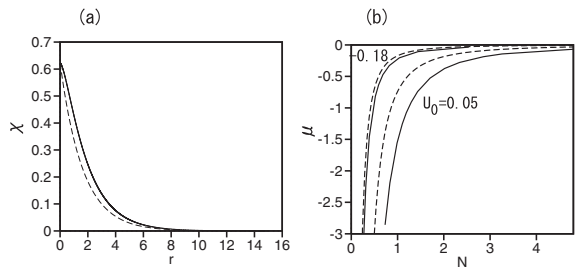}
\end{center}
\caption{(a) The radial profile of the ground state in the 2D model with the
quintic nonlinearity and $\Omega =0$, for $U_{0}=0.05$ and $\protect\mu %
=-0.1867$. (b) Curves $\protect\mu (N)$ for the GS families with $U_{0}=-0.18
$ and $U_{0}=0.05$. In both panels, the numerical results and the respective
analytical approximation (\protect\ref{inter2D}) are depicted by the
continuous and dashed curves. The convergence of the numerical and
analytical curves for $N(\protect\mu )$ at $\protect\mu \rightarrow -0$
corresponds to the fact that Eq. (\protect\ref{inter2D}) gives an
asymptotically exact solution in this limit.}
\label{fig5}
\end{figure}

An example of the stable GS, and curves $\mu (N)$ for the GS families in 2D
are displayed, along with the analytical approximation (\ref{inter2D}), in
Fig. \ref{fig5}. The $\mu (N)$ curves are shown for both signs of the
central potential (and $l=0$), $U_{0}=-0.18$ and $U_{0}=0.05$. Simulations
of perturbed solutions within the framework of the radial version of Eq. (%
\ref{psi2d}) confirm the stability of the GS families (not shown here).
Although these simulations do not include azimuthal perturbations, the
repulsive sign of the nonlinearity makes it evident that these perturbations
will not give rise to an instability \cite{review}.

\subsection{Effects of the harmonic trap}

\begin{figure}[t]
\begin{center}
\includegraphics[height=3.5cm]{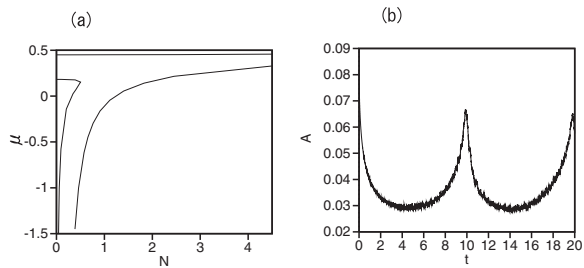}
\end{center}
\caption{(a) Curves $\protect\mu (N)$ for three stationary modes found in
the 2D model with the quintic nonlinearity for $U_{0}=-0.18$ and $\Omega
^{2}=0.10$, cf. the similar picture for the 3D cubic model displayed in Fig.
\protect\ref{fig3}. (b) An example of the breather generated by the solution
with $N=0.3453$ belonging to the unstable middle branch from panel (b). The
plot shows $A\equiv \left\vert \protect\chi (r=0,t)\right\vert ^{2}\ $as a
function of time, with $\protect\chi $ defined as $\protect\chi _{\mathrm{2D}%
}$ in Eq. (\protect\ref{psichi2D}).}
\label{fig6}
\end{figure}

As shown in Fig. \ref{fig6}(a), in the presence of the external trap ($%
\Omega >0$), three different stationary solutions can be found at $%
-1/4<U_{l}<0$, when the corresponding 2D linear Schr\"{o}dinger equation
gives rise to two exact solutions, in the form given by Eqs. (\ref{2D}), (%
\ref{exact}), and (\ref{sigma2D}). As well as in the similar situation for
the 3D case, which is displayed in Fig. \ref{fig3}(a), two upper branches in
Fig. \ref{fig6} represent nonlinear deformations of the exact linear
solutions, with values of $\mu $ approaching those given by Eq. (\ref%
{sigma2D}) in the limit of $N\rightarrow 0$. The lowest branch represents
the GS corresponding to approximation (\ref{inter2D}), which is additionally
deformed by the trapping potential. In the limit of $\mu \rightarrow \infty $%
, both the bottom and top branches in Fig. \ref{fig6}(a) asymptotically
approach the Thomas-Fermi limit, $\left( N_{\mathrm{TF}}\right) _{\mathrm{2D}%
}\approx \left( 4\pi /3\Omega ^{2}\right) \mu ^{3/2}$, cf. Eq. (\ref{TF3D}).

In spite of the overall similarity to the 3D case, it is worthy to note that
the shape of the middle branch in Fig. \ref{fig6}(a) is drastically
different from its counterpart in the 3D case, cf. \ref{fig3}(a). Moreover,
direct simulations of Eq. (\ref{psi2d}) demonstrate that, while the
solutions corresponding to the top and bottom branches in Fig. \ref{fig6}(a)
are stable (not shown here in detail), the middle-branch solutions are \emph{%
not}. In the simulations, they are spontaneously transformed into robust
breathers featuring long-period oscillations, see an example in Fig. \ref%
{fig6}(b).

\section{Conclusion}

We have demonstrated that the mean-field repulsive nonlinearity suppresses
the 3D quantum collapse induced by the central attractive potential, $%
-\left( U_{0}/2\right) r^{-2}$, which can be realized in the ultracold gas
of dipolar molecules attracted by the central charge. The dipole-dipole
interactions were also taken into account, resulting (in the framework of
the mean-field approximation) in a redefinition of the scattering length
which accounts for the contact repulsion. The nonlinearity creates the GS
(ground state) in place of the collapse regime. For $U_{0}<1/4$, when the
respective 3D Schr\"{o}dinger equation does not lead to the collapse, the
inclusion of the harmonic trap gives rise to the tristability. The cubic
repulsion is not strong enough to prevent the 2D collapse, but the quintic
term is sufficient for this purpose. It also gives rise to the GS which
replaces the quantum collapse in the 2D space, and to similar modes carrying
the angular momentum (which feature the amplitude diverging, rather than
vanishing, at $r\rightarrow 0$, while the total norm of the vortical mode
converges). A difference from the 3D case is that, in the presence of the
harmonic trap, one of the three confined modes supported by the weakly
repulsive central potential with $0<-U_{0}<1/4$, in the combination withe
quintic nonlinearity, is unstable, transforming itself into a breather. A
counter-intuitive finding is that the 2D self-trapped mode exists even in
the case of the weakly repulsive potential, while the harmonic trap is
absent.

This work suggests continuations in several directions. If the orientation
of the dipoles in the 3D space is fixed by an external uniform field, the
central charge induces an axisymmetric potential, $U=-\left( U_{0}/2\right)
r^{-2}\cos \theta $, and it may be interesting to study the possibility of
the replacement of the corresponding anisotropic collapse (cf. Ref. \cite%
{anomaly2}) by a GS. It is relevant to mention that the dipole-dipole
interactions between the bosons may give rise to a specific mode of the
nonlinear collapse in BEC \cite{d-wave}, in the absence of the contact
repulsion, which suggests to consider an interplay of this nonlinear mode
with the linear collapse (although the dynamics of the collapse is not
described by the mean-field approximation \cite{Pit}). Another interesting
extension may be the study of higher-order nonlinear states, such as 3D
vortical modes.

A challenging problem is to extend the analysis to fermion gases. In that
connection, it is relevant to mention that the fall-onto-the-center effect
was studied for interacting fermion pairs, in the framework of the
Bethe-Salpeter equation \cite{Efimov}.

\end{document}